\def\cm{{\rm\thinspace cm}}

\def\km{{\rm\thinspace km}}

\def\Msun{\hbox{$\rm\thinspace M_{\odot}$}}

\def\s{{\rm\thinspace s}}
\def\yr{{\rm\thinspace yr}}

\def\kmps{\hbox{$\km\s^{-1}\,$}}

\def\Msunpyr{\hbox{$\Msun\yr^{-1}\,$}}
\def\pcm{\hbox{$\cm^{-2}\,$}}

\documentclass{iaus}
\usepackage{graphicx}

\title[Cosmic Feedback from AGN] 
{Cosmic Feedback from AGN}

\author[A.C Fabian]   
{A.C. Fabian}

\affiliation{Institute of Astronomy, Madingley Road, Cambridge CB3
  0HA, UK \\ email: {\tt acf@ast.cam.ac.uk} }

\pubyear{2009}
\volume{267}  
\pagerange{119--126}
\jname{Co-Evolution of Central Black Holes and Galaxies}
\editors{B.M.\ Peterson, R.S.\ Somerville, \& T.\ Storchi-Bergmann, eds.}
\begin{document}

\maketitle

\begin{abstract}
  Accretion onto the massive black hole at the centre of a galaxy can
  feed energy and momentum into its surroundings via radiation, winds
  and jets. Feedback due to radiation pressure can lock the mass of
  the black hole onto the $M_{\rm BH}-\sigma$ relation, and shape the
  final stellar bulge of the galaxy. Feedback due to the kinetic power
  of jets can prevent massive galaxies greatly increasing their
  stellar mass, by heating gas which would otherwise cool radiatively.
  The mechanisms involved in cosmic feedback are discussed and
  illustrated with observations.  
\end{abstract}

\firstsection 
\section{Introduction}
It has been realised over the past decade that the black hole at the
centre of a galaxy bulge is no mere ornament but plays a major role in
determining the final stellar mass of the galaxy. The process by which
this occurs is known as cosmic feedback and it takes place through an
interaction between the energy and radiation generated by accretion
onto the massive black hole and the gas in the host galaxy. The
ratio of the size of the black hole to that of the galaxy is huge and 
similar to a person in comparison to the Earth, so the details of the
feedback process are complex. 

The overall picture in terms of energetics is fairly straightforward
and two major modes have been identified. The first is the {\it
  radiative} mode, also known as the quasar or wind mode, which
operates, or operated, in a typical bulge when the accreting black
hole was close to the Eddington limit. The second mode is the {\it
kinetic} mode, also known as the radio or jet mode. This typically
operates when the galaxy has a hot halo (or is at the centre of a
group or cluster of galaxies) and the accreting black hole has
powerful jets. It tends to occur at a lower Eddington fraction and
in the more massive galaxies. 

It is easy to demonstrate that the growth of the central black hole by
accretion can have a profound effect on its host galaxy. If the
velocity dispersion of the galaxy is $\sigma$ then the binding energy
of the galaxy, which is of mass $M_{\rm gal},$ is $E_{\rm gal}\approx
M_{\rm gal} \sigma^2$. The mass of the black hole $ M_{\rm BH}\approx
2\times 10^{-3} M_{\rm gal}$ (Tremaine et al 2002, H\"aring \& Rix
2004). Assuming a radiative efficiency for the accretion process of
10\%, then the energy released by the growth of the black hole $E_{\rm
  BH}= 0.1 M_{\rm BH} c^2$. Therefore $E_{\rm BH}/E_{\rm gal}\approx
2\times 10^{-4}(c/\sigma)^2.$ For a galaxy $\sigma<450\kmps,$ 
so $E_{\rm BH}/E_{\rm gal} > 100$.

Fortunately accretion energy does not significantly affect the stars
of the host galaxy, or there would not be any galaxies. The energy and
momentum from accretion do couple with the gas. The processes involved
in that are reviewed here.

\section{The Radiative or Wind Mode}
Silk \& Rees (1998, see also Haehnelt et al 1998) point out that a
quasar at the Eddington limit can prevent accretion into a galaxy at
the maximum possible rate (equivalent to its gas content going into
free fall at a rate $\sim f\sigma^3/G$, so power needed is $\sim
f\sigma^5/G$) provided that
$$M_{\rm BH} \sim{{f\sigma^5\sigma_{\rm T}}\over{G^2 m_{\rm p} c}},$$
where $\sigma_{\rm T}$ is the Thomson cross section for electron
scattering and $f$ is the fraction of the galaxy mass in gas . The
galaxy is assumed to be isothermal with radius $r$, so that its mass
is $2\sigma^2 r/G.$ The argument is based on energy balance.

Momentum balance gives an expression (Fabian 1999, Fabian et al 2002,
King 2003, 2005, Murray et al 2005)
$$M_{\rm BH}={{f\sigma^4\sigma_{\rm T}}\over{\pi G^2 m_{\rm p}}},$$
which is about $c/\sigma$ times larger and more in line with the
observed $M_{\rm BH}-\sigma$ relation for $f\sim 0.1$.

There are several ways to derive the above formula. A simple one is to
assume that the radiation pressure from the Eddington-limited quasar
has swept the gas, of mass $fM_{\rm gal},$ to the edge of the
galaxy. Balancing forces gives
$${L_{\rm Edd}\over c}={{GM_{\rm gal}M_{\rm gas}}\over r^2}$$
i.e.
$${{4\pi G M_{\rm BH}m_{\rm p} }\over{\sigma_{\rm T} }}={Gf}\left({{2\sigma^2 }\over G}\right)^2,$$
from which the result follows.

The interaction cannot rely on radiation pressure on electrons as in
the standard Eddington-limit formula, since if the quasar is locally
at its Eddington limit then it must be far below the Eddington limit
when the mass of the galaxy is included. (Quasars appear to respect the
Eddington limit, see e.g. Kollmeier 2006.) The interaction has to be
much stronger, either through a wind generated close to the quasar
which then flows through the galaxy pushing the gas out, or to dust in
the gas, as expected for the interstellar medium of a galaxy (Laor \&
Draine 1993, Scoville \& Norman 1995, Murray et al 2005). Dust
grains embedded in the gas will be partially charged in the energetic
environment of a quasar which will link them to the surrounding
partially-ionized gas. $L_{\rm Edd}$ is reduced by a factor of
$\sigma_{\rm d}/\sigma_{\rm T}$, where $\sigma_{\rm d}$ is the
equivalent dust cross section per proton, appropriately weighted for
the dust content of the gas and the spectrum of the quasar.

\begin{figure}[h]
\begin{center}
 \includegraphics[width=2.8in]{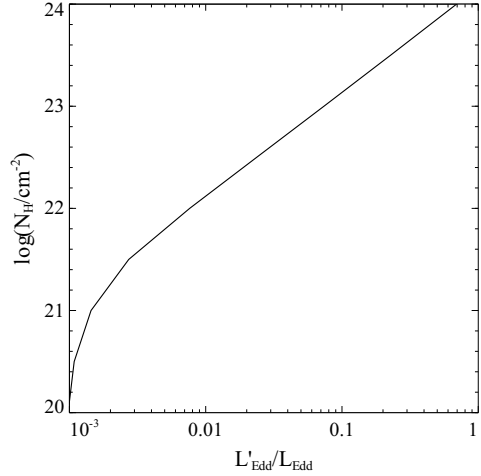} 
 \caption{Column density $N_{\rm H}$ of dusty gas required for a given
   effective Eddington ratio $L_{\rm Edd}'/L_{\rm
     Edd}$. A quasar
   spectrum and Galactic dust fraction is assumed, giving $\sigma_{\rm
     d}/\sigma_{\rm T}\sim 1000.$ The result for a typical AGN
   operating at $L/L_{\rm Edd}<0.1$ is $\sim 300$. A value of 500 is
   adopted in the text. }
   \label{fig1}
\end{center}
\end{figure}

We find that $\sigma_{\rm d}/\sigma_{\rm T}$ is about 500 for a
Galactic dust-to-gas ratio (Fabian et al 2008). This means that a quasar at
the standard Eddington limit (for ionized gas) is at the effective
Eddington limit (for dusty gas), $L_{\rm Edd}',$ of a surrounding
object 500 times more massive (Fig.~1). Is this just a coincidence or the
underlying reason why $ M_{\rm gal}/M_{\rm BH}\approx 500$?

\begin{figure}[b]
\begin{center}
 \includegraphics[width=2.5in]{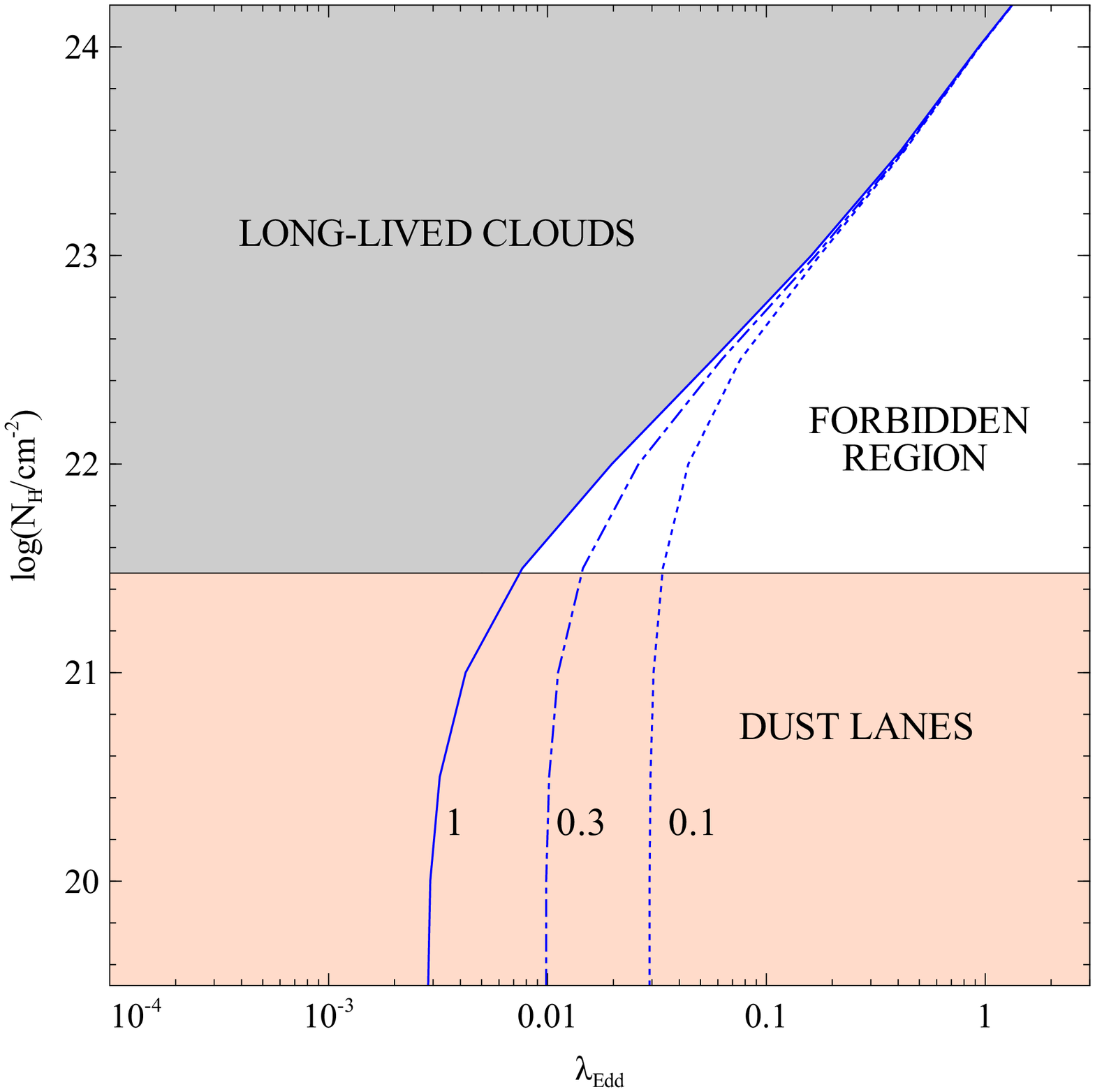} 
\hspace{0.5cm}
\includegraphics[width=2.5in]{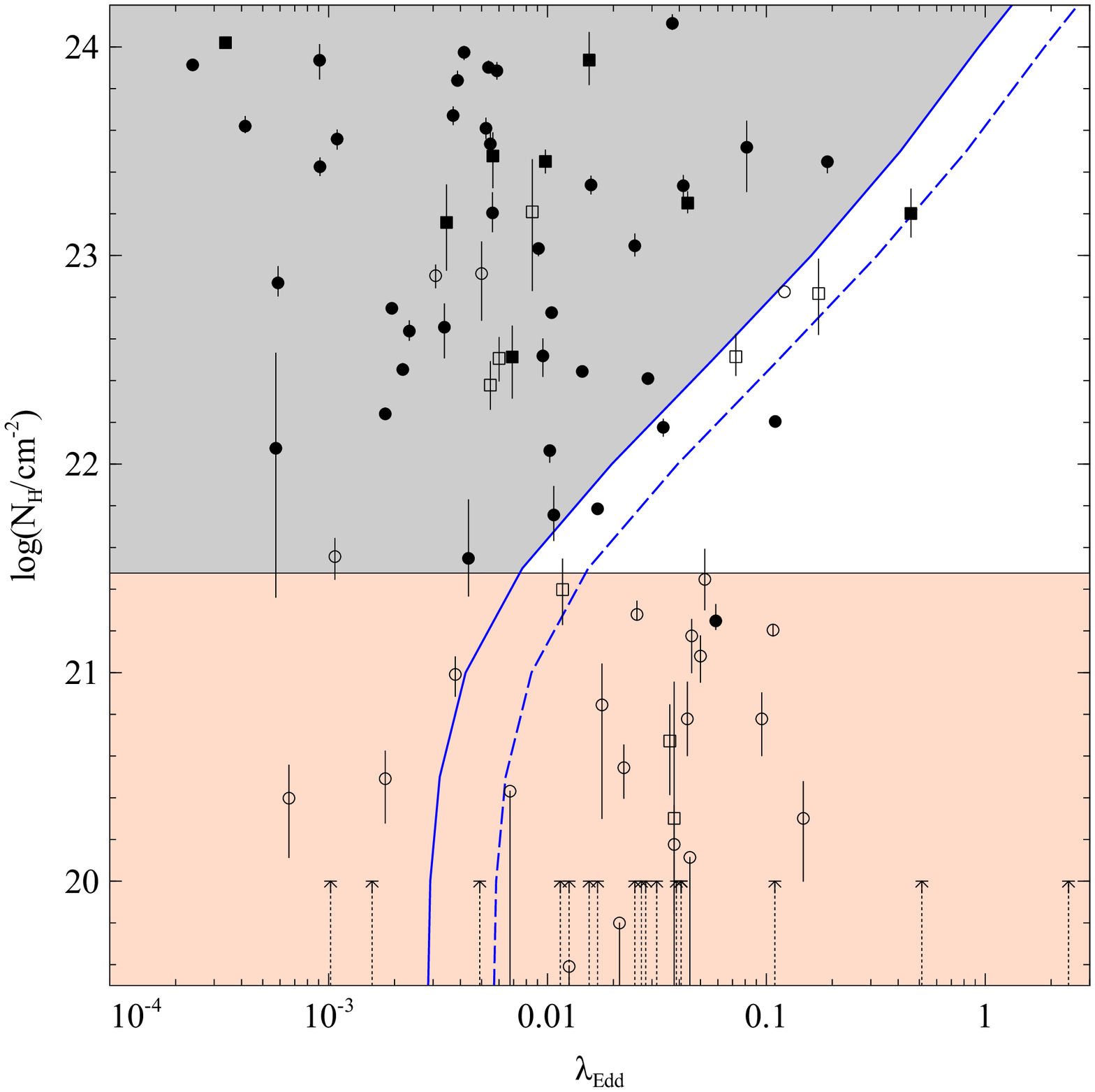}
\caption{The forbidden region in the ($\lambda, N_{\rm H}$)
  plane. Dust lanes can occur at large radii at low columns. The 3
  curves indicate different dust to gas ratios relative to the
  Galactic value.  On the right is the plane populated by the
  SWIFT-BAT AGN. The dashed curve is for $L_{\rm Edd}$ appropriate for
  a region of twice the black hole mass expected at a radius of few pc
  (Fabian et al 2009). The single point in the forbidden region is a
  galaxy with an outflowing warm absorber.}
   \label{fig1}
\end{center}
\end{figure}

We have investigated (Fabian et al 2006, 2008, 2009) whether there are
indications of $L_{\rm Edd}'$ in AGN by examining the plane of
absorption column density, $N_{\rm H}$ versus Eddington fraction,
$\lambda=L_{\rm bol}/L_{\rm Edd}$. Most intrinsic cold absorption seen
in X-ray spectra originates fairly close to the nucleus (the required
gas mass would otherwise be prohibitive). Therefore an AGN with
$\lambda=1/500$ is highly sub-Eddington for local ionized dust-free
gas, but effectively at the Eddington limit for dusty gas clouds which
are optically thin to dust absorption. The outer parts of larger
optically-thick clouds act as dead weight which increases $L_{\rm
  Edd}'$ (Fig.~1). The net result is that there should be a forbidden
zone in the ($\lambda, N_{\rm H}$) plane to the right of the $L_{\rm
  Edd}'$ curve. This is indeed what we see (Fig.~2). At low column
densities there can be absorption from an outer disc or dust lane, so
we ignore the region below about $3\times 10^{21}\pcm$. The observed
points show that AGN do indeed influence the amount of gas in a galaxy
bulge.

The way that gas can evolve on the ($\lambda, N_{\rm H}$) plane is
that gas which strays into the forbidden zone is pushed outward,
reducing $ N_{\rm H}$ as it does so in a shell, sliding along the
curve. Gas which is introduced to a galaxy can stay, fuelling both the
black hole and star formation, provided both $L_{\rm Edd}'$ and
$L_{\rm Edd}$ remain below unity. Repitition of this process should
drive $ M_{\rm BH}/ M_{\rm gal}\rightarrow 2\times 10^{-3}$. At higher
redshifts where the metallicity and dust content was less, then this
ratio should be proportionately higher. Studies of the ($\lambda,
N_{\rm H}$) plane at higher redshifts and with large samples may show
us how the gas content of galaxies responds and evolves to the action
of quasars.

If the repeated action of
radiation pressure on dust is responsible for the $M_{\rm BH}-\sigma$
relation then it must cause the bulge mass
$${M_{\rm gal}}\sim {{f\sigma^4\sigma_{\rm d}}\over{\pi G^2
    m_{\rm p}}}.$$
or $${{\sigma^2}\over r}\sim {{2\pi G m_{\rm p}}\over {f\sigma_{\rm
      d}}}.$$
Feedback should shape both the black hole and the galaxy bulge and may even
lead to the fundamental plane.

If the main interaction is due to winds, not to radiation pressure,
then the wind needs to have a high column density, high
velocity $v$, high covering fraction $f$, all at large radius $r$. The
kinetic luminosity of a wind is 
$${{L_{\rm w}}\over L_{\rm Edd}}={f\over 2}{r\over r_{\rm
    g}}\left({v\over c}\right)^3{N\over N_{\rm T}},$$ where $r_{\rm
  g}$ is the gravitational radius $GM/c^2$ and $N_{\rm T}=\sigma_{\rm
  T}^{-1}.$ To produce $M_{\rm BH}-\sigma^4$ scaling the thrust of the
wind needs to correspond to the Eddington limit (the wind may need to
be dusty).  Warm absorbers flowing at $\sim 1000\kmps$ are
insufficient (Blustin et al 2005). BAL quasars flowing at tens of
thousands $\kmps$ may however be important (see article in these
Proceedings by N. Arav).
 
\section{The Kinetic Mode}

The more massive galaxies at the centres of groups and clusters are
often surrounded by gas with a radiative cooling time short enough
that a cooling flow should be taking place. The mass cooling rates
would be tens, hundreds or even thousands of $\Msunpyr$. Such objects
should be significantly growing their stellar mass now, yet they are
not. This is because the massive black hole at the centre of the
galaxy is feeding energy back into its surroundings at a rate
balancing the loss of energy through cooling (for reviews see Peterson
\& Fabian 2006, McNamara \& Nulsen 2007, Cattaneo et al 2009).

\begin{figure}[h]
\begin{center}
\includegraphics[width=3in]{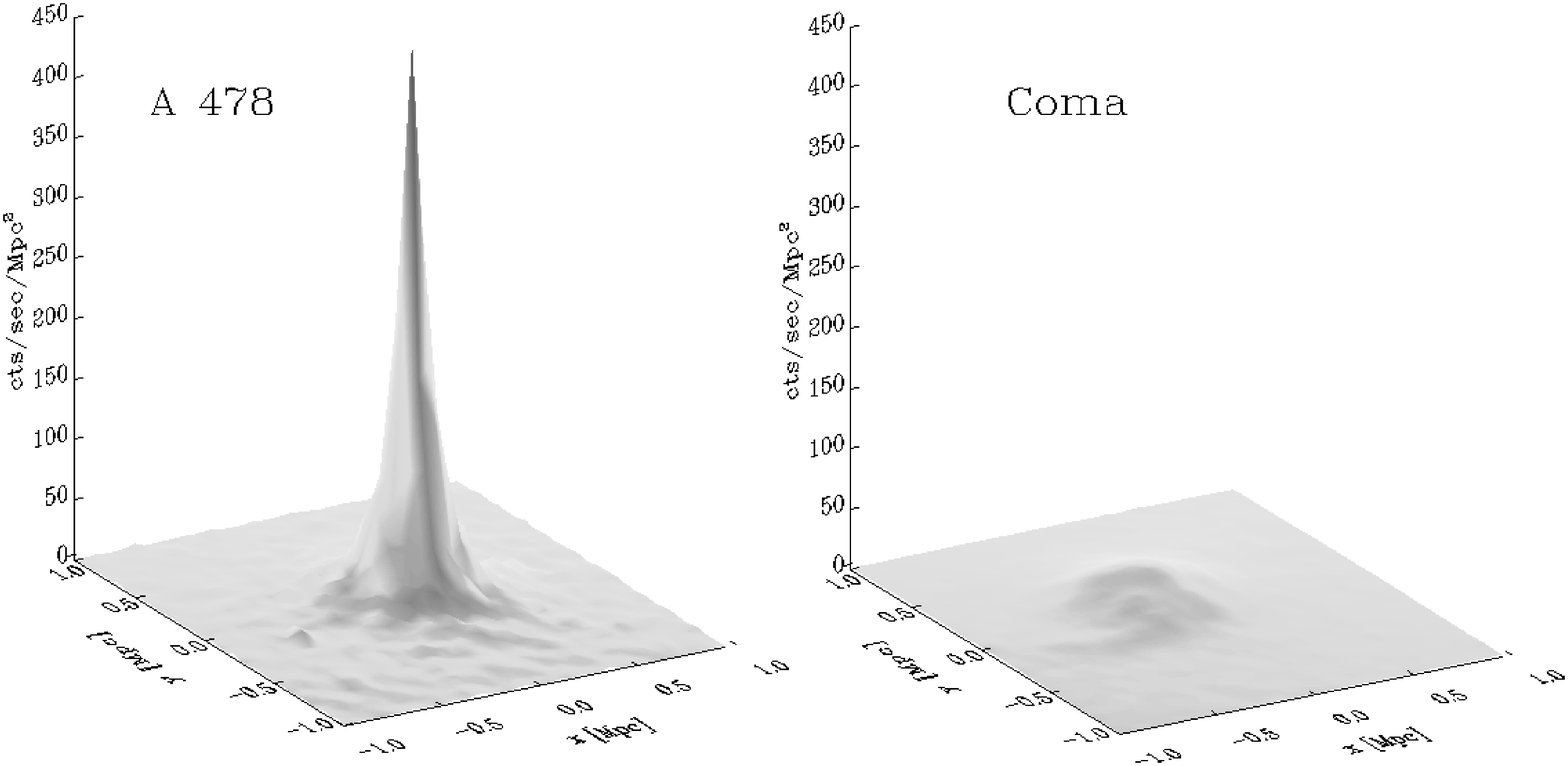} 
\includegraphics[width=2.2in]{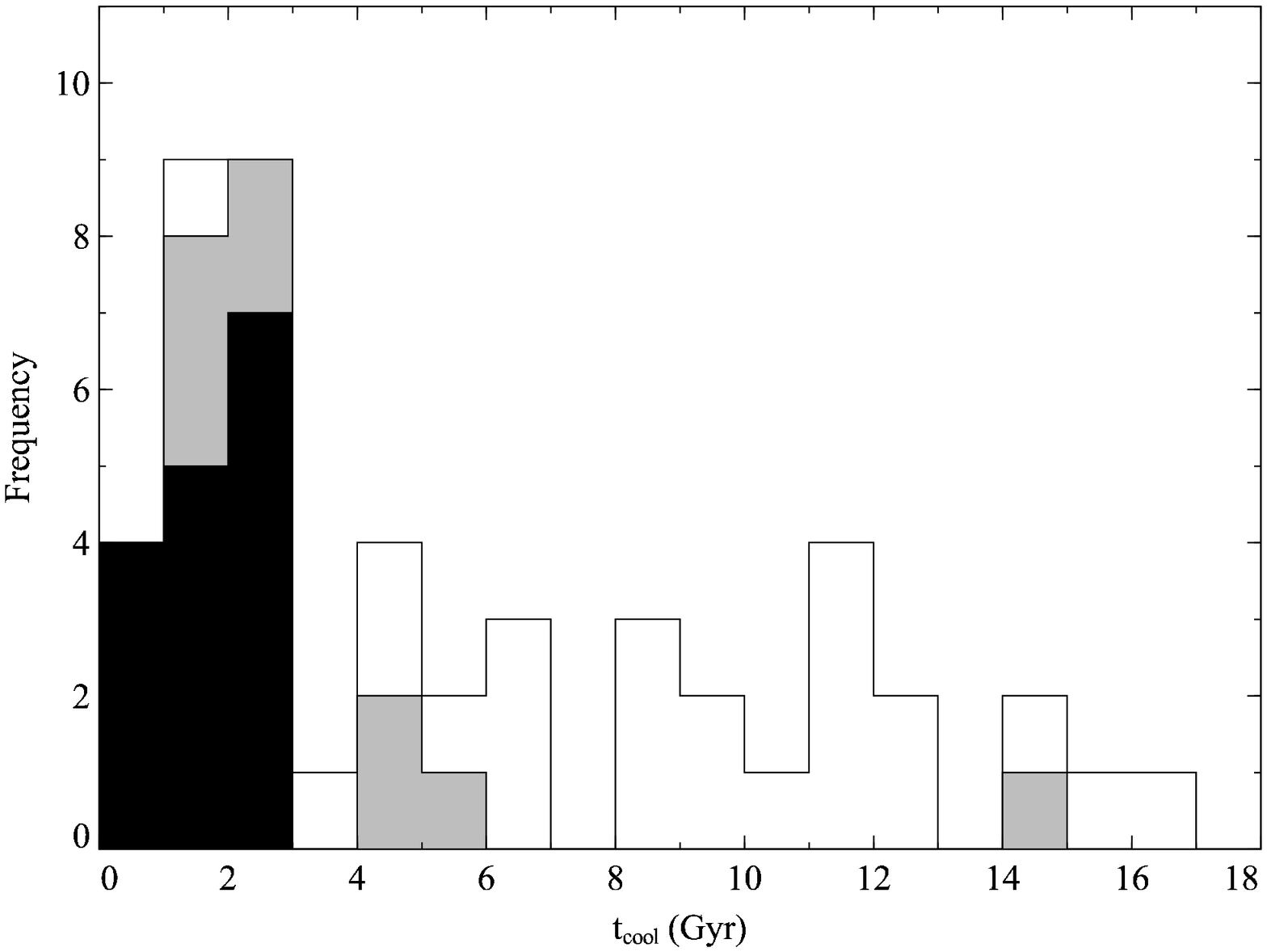} 
\caption{Left: The X-ray surface brightness peak at the centre of a
  cool core cluster, A478, which has a short central cooling time,
  compared with a cluster with a longer cooling time, Coma. Histogram
  of cooling times in the B55 cluster sample (Dunn \& Fabian
  2006). Black indicates bubbles seen and grey that there is a central
  radio source.}
   \label{fig1}
\end{center}
\end{figure}

Several steps in this feedback process are clearly seen in X-ray and
radio observations. The accretion flow onto the black hole generates
powerful jets which inflate bubbles of relativistic plasma either side
of the nucleus. The bubbles are buoyant in the intracluster or
intragroup medium, separating and rising as a new bubble forms, if the
jet operates more or less continuously (Churazov et al 2001). This is
seen to be the case since a study of the brightest 55 clusters
(Fig.~4; Dunn et al 2006; see also Rafferty et al 2006) shows that over
70\% of those clusters where the cooling time is less than 3~Gyr,
which need heat, have bubbles, and another 20\% have a central radio
source.

\begin{figure}[h]
\begin{center}
 \includegraphics[width=2.6in]{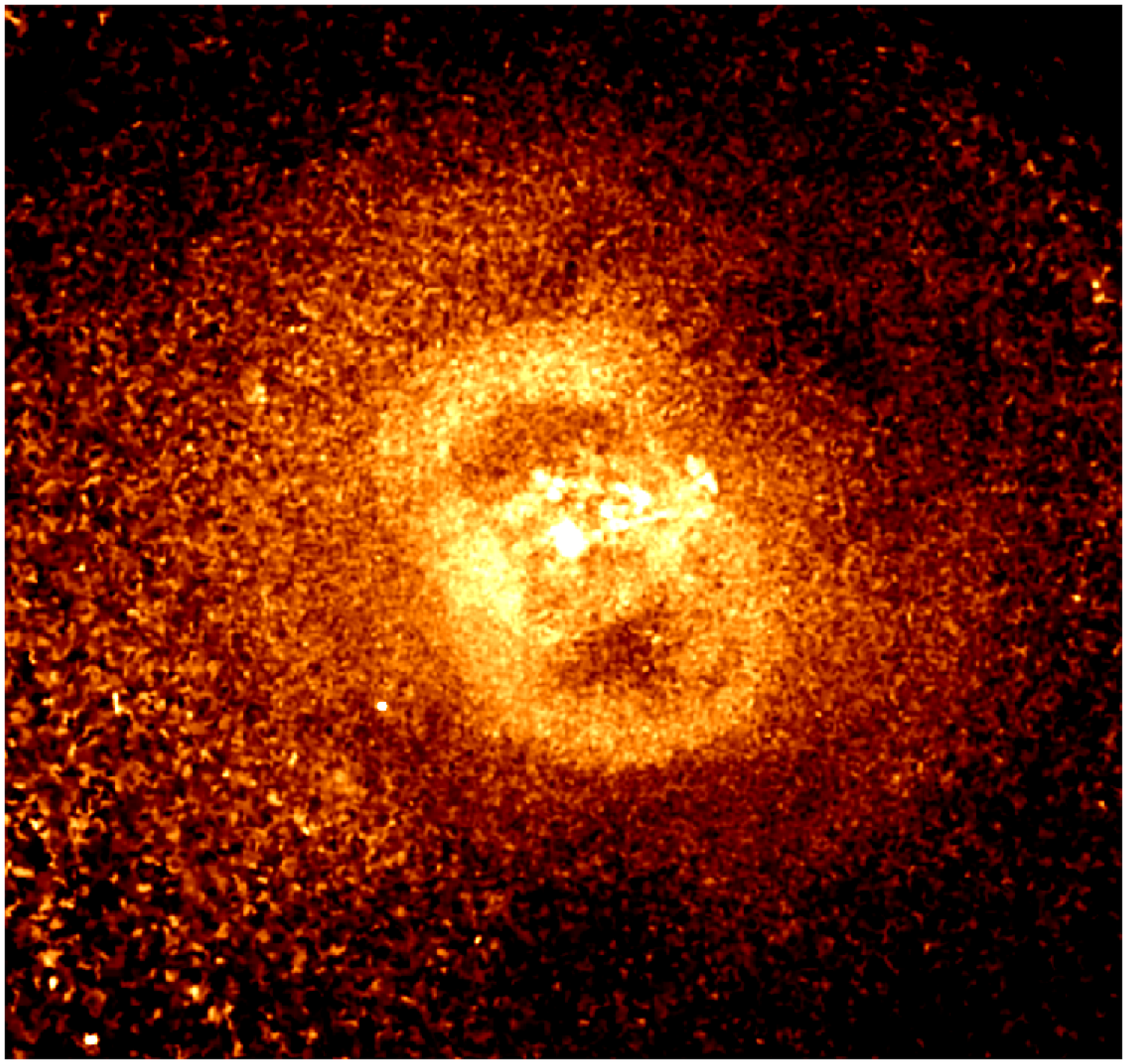} 
\hspace{0.3cm}
\includegraphics[width=2.45in]{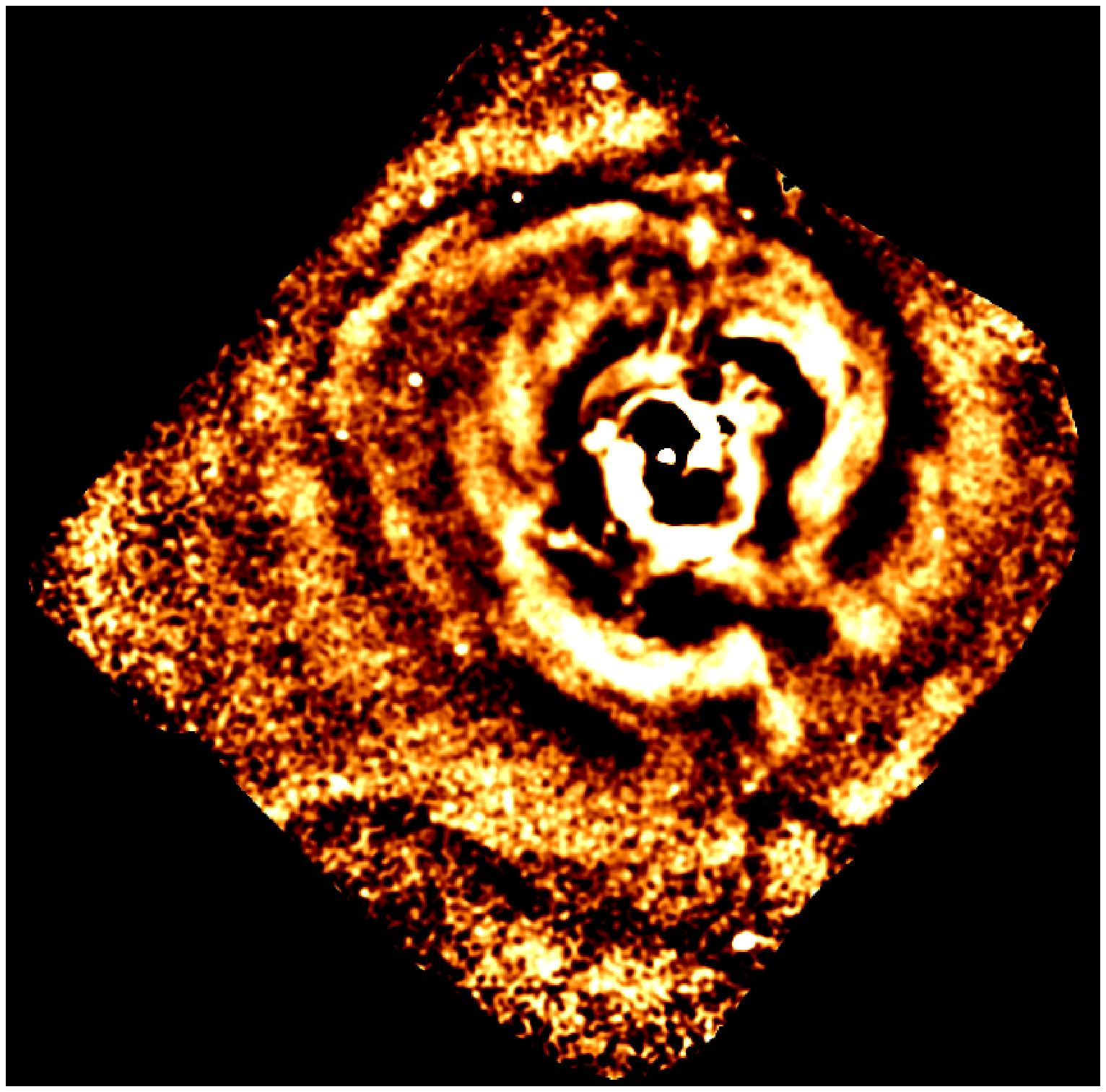}
\caption{Left: Pressure map derived from Chandra imaging X-ray
  spectrsocopy of the Perseus cluster. Note the thick high pressure
  regions containing almost $4PV$ of energy surrounding each inner
  bubble, where $V$ is the volume of the radio-plasma filled interior
  (Fabian et al 2006). Right: unsharp masked image showing the
  pressure ripples or sound waves. }
   \label{fig1}
\end{center}
\end{figure}

The kinetic power in the jets can be estimated from the size of the
bubbles, the surrounding pressure (obtained from the density and
temperature of the thermal gas) and the buoyancy time (which depends
on the gravitational potential). The power is high and only weakly
correlated with radio power (the radiative efficiency of many jets is
very low at between $10^{-2}-10^{-4}$). The power is usually in good
agreement with the energy loss by X-radiation from the
short-cooling-time region.  The overall energetics of the feedback
process are therefore not an issue.

In the case of the Perseus cluster, which is the X-ray brightest in
the Sky, Chandra imaging shows concentric ripples which we
interpret as sound waves generated by the expansion of the central
pressure peaks associated with the repetitive blowing of bubbles
(Fabian et al 2003, 2006). The energy flux in the sound waves is
comparable to that required to offset cooling, showing that this is
the likely way in which heat is distributed in a quasi-spherical
manner. Similar sound waves, or weak shocks, are also seen in the
Virgo, Centaurus and A2052 clusters and in simulations (Ruszkowski et
al 2004; Sijacki \& Springel 2006).

Let us now consider how the apparent close heating/cooling balance has
been established and maintained. The lack of high star formation rates
suggests that cooling does not exceed heating by ten  per cent or
so. The presence of central abundance gradients and pronounced
temperature drops indicates that heating does not general exceed
cooling by that much either. This balance needs to continue over tens
to hundreds of bubbling cycles (each of a 10--50~Myr or so).

\begin{figure}[h]
\begin{center}
 \includegraphics[width=2.8in]{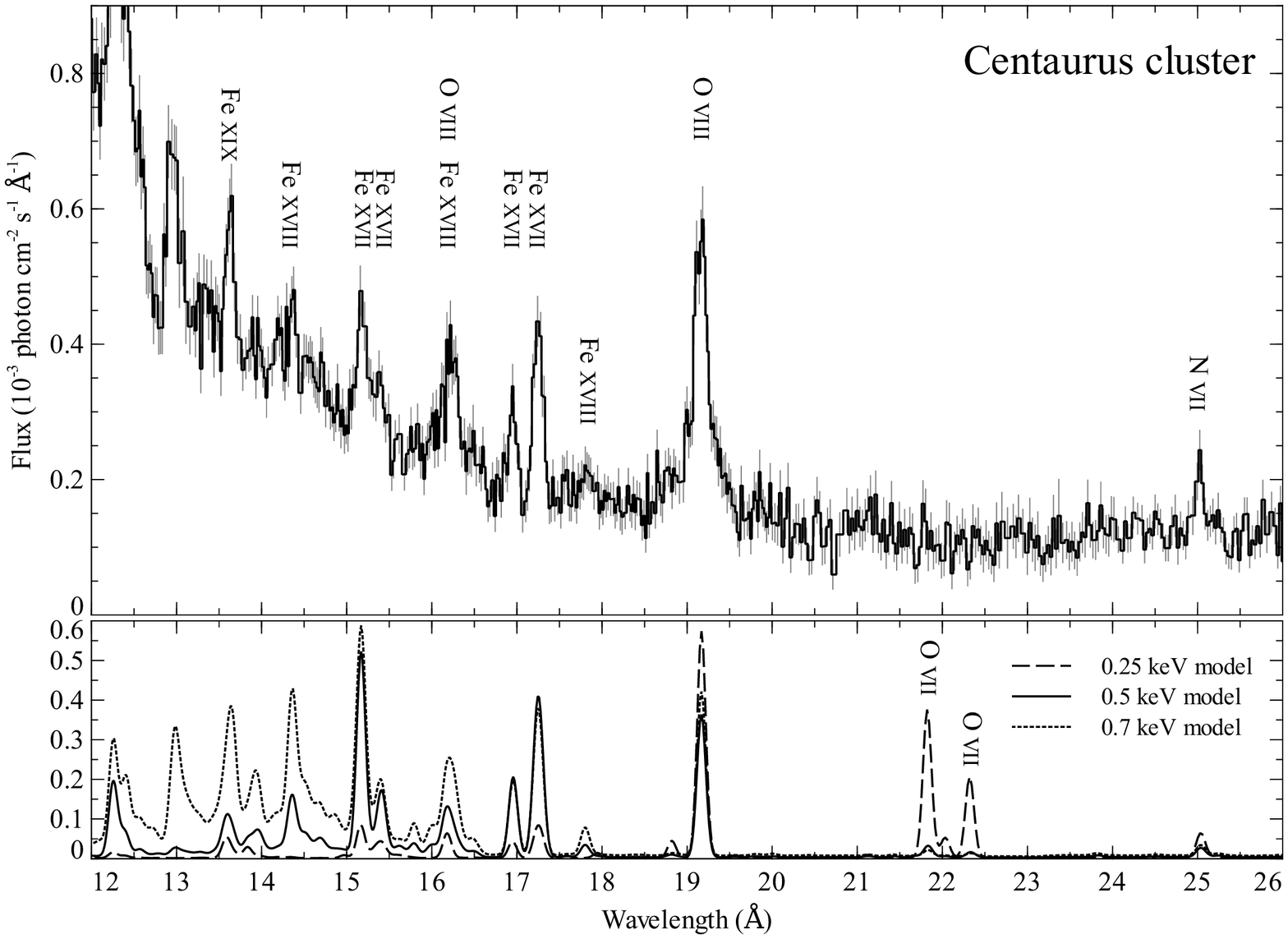} 
\hspace{0.5cm}
\includegraphics[width=2.05in]{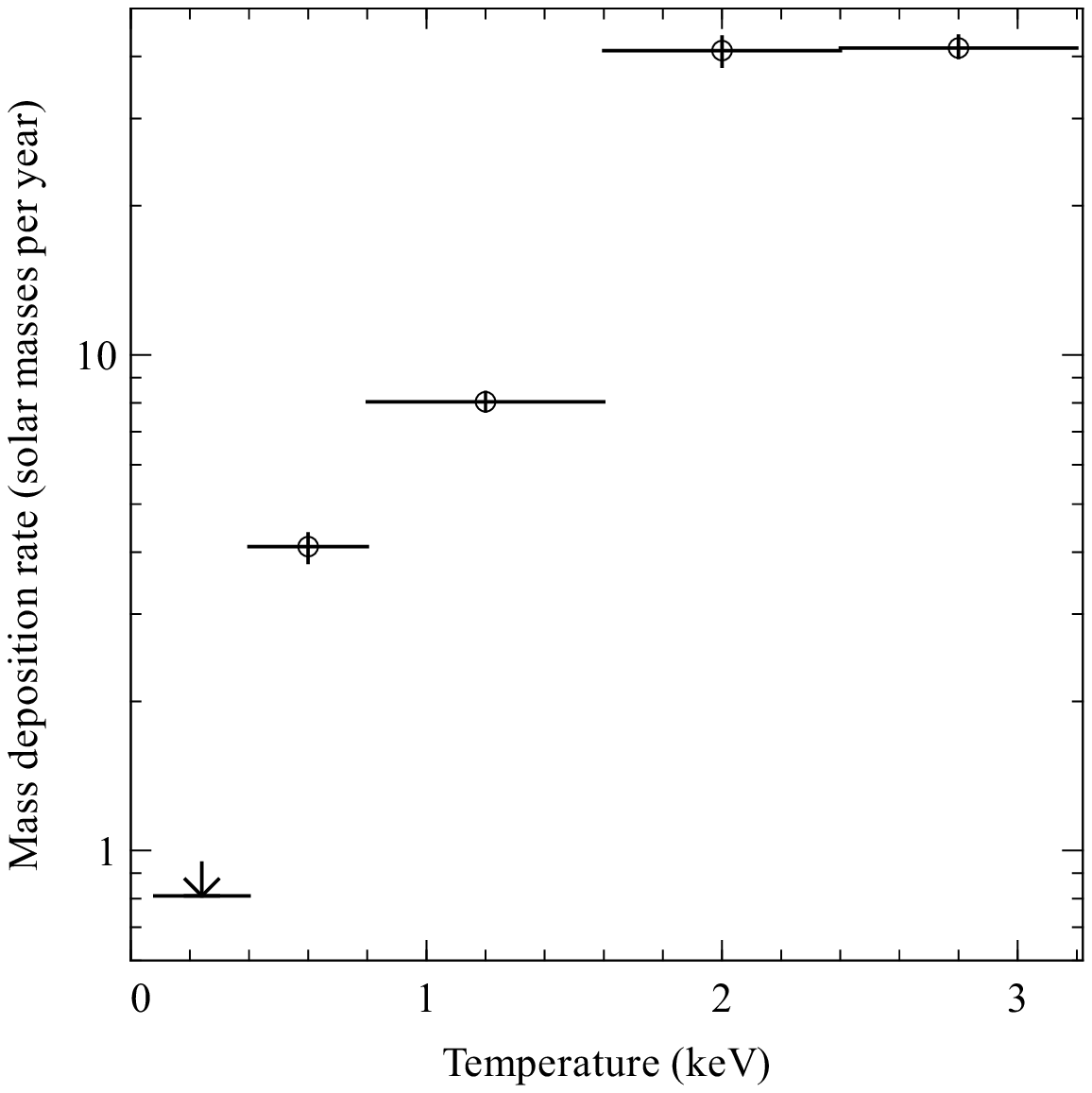}
\caption{Left: RGS spectrum of the centre of the nearby Centaurus
  cluster showing strong FeXVII and OVIII lines but no OVII (Sanders
  et al 2007). Right: Mass cooling rate of gas. Note that little gas
  seems to cool below 0.5~keV.  }
   \label{fig1}
\end{center}
\end{figure}

A simple 1D feedback cycle seems at first sight possible. If too much
gas starts to cool then the accretion rate should increase making the
heating rate go up and vice versa.  However the time scales involved are
long, meaning that feedback would be delayed and angular momentum
could prevent gas reaching anywhere near the black hole.

Studying the details requires the best data on the brightest nearby
objects.  X-ray images from Chandra and moderate resolution spectra
from the XMM-Newton Reflection Grating Spectrometer (RGS) show X-ray
cool gas in some clusters cores, with temperatures ranging from 5 to
0.5~keV in the nearby Centaurus cluster (Sanders et al 2007).  The
coolest gas has a cooling time of only 10~Myr, yet the spectra show no
sign of any lower gas temperature gas (where OVII emission is
expected). In this object the heating/cooling balance looks to hold to a
few per cent. How the 0.5~keV gas is prevented from cooling is not
obvious. The images show that it is clumpy so the question arises as
to how it is targetted for heating without its immediate surroundings
being overheated. A similar picture emerges from several other
clusters with excellent data.

\begin{figure}[h]
\begin{center}
 \includegraphics[width=2.5in]{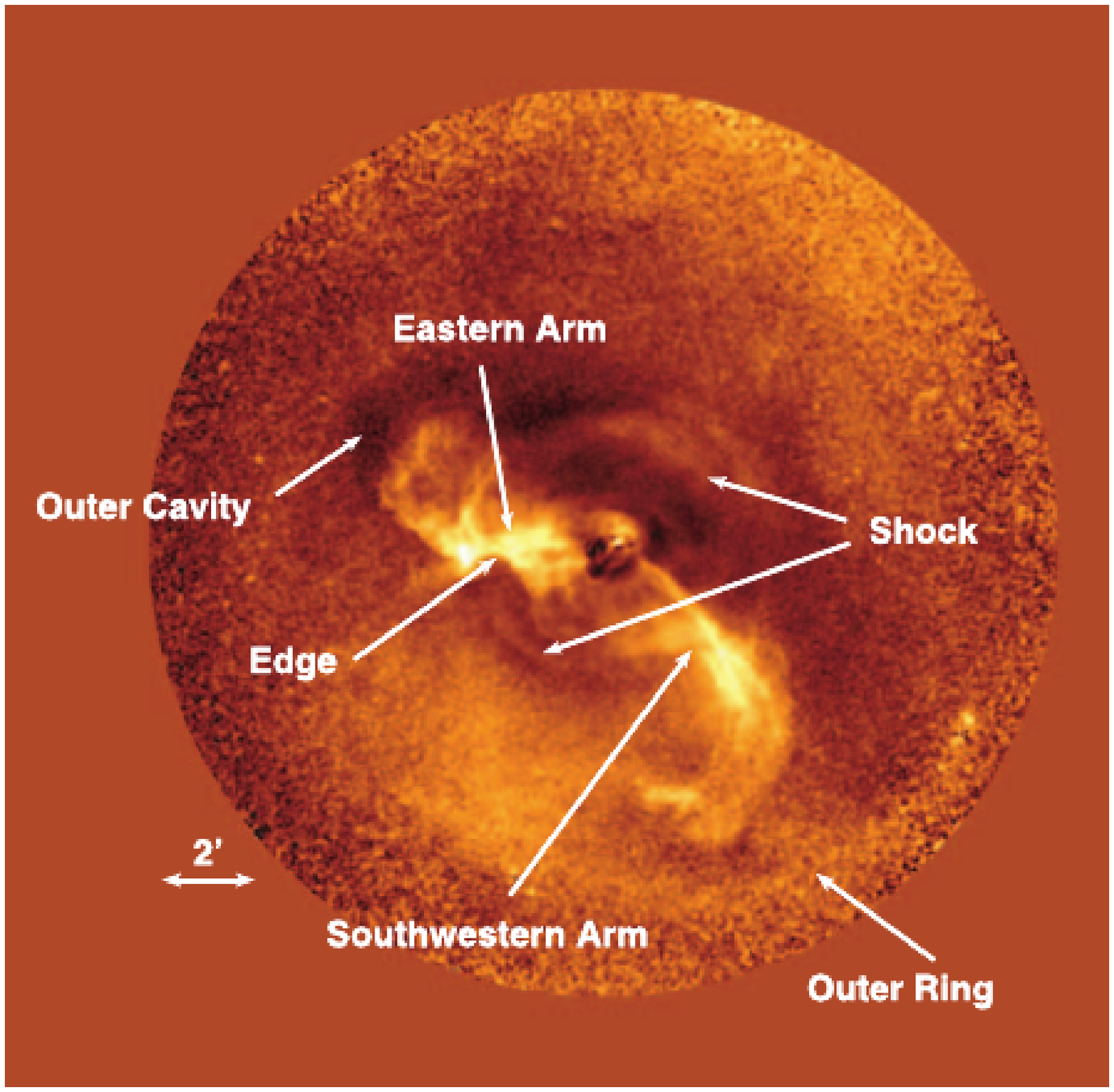}
\hspace{0.3cm}
\includegraphics[width=2.5in]{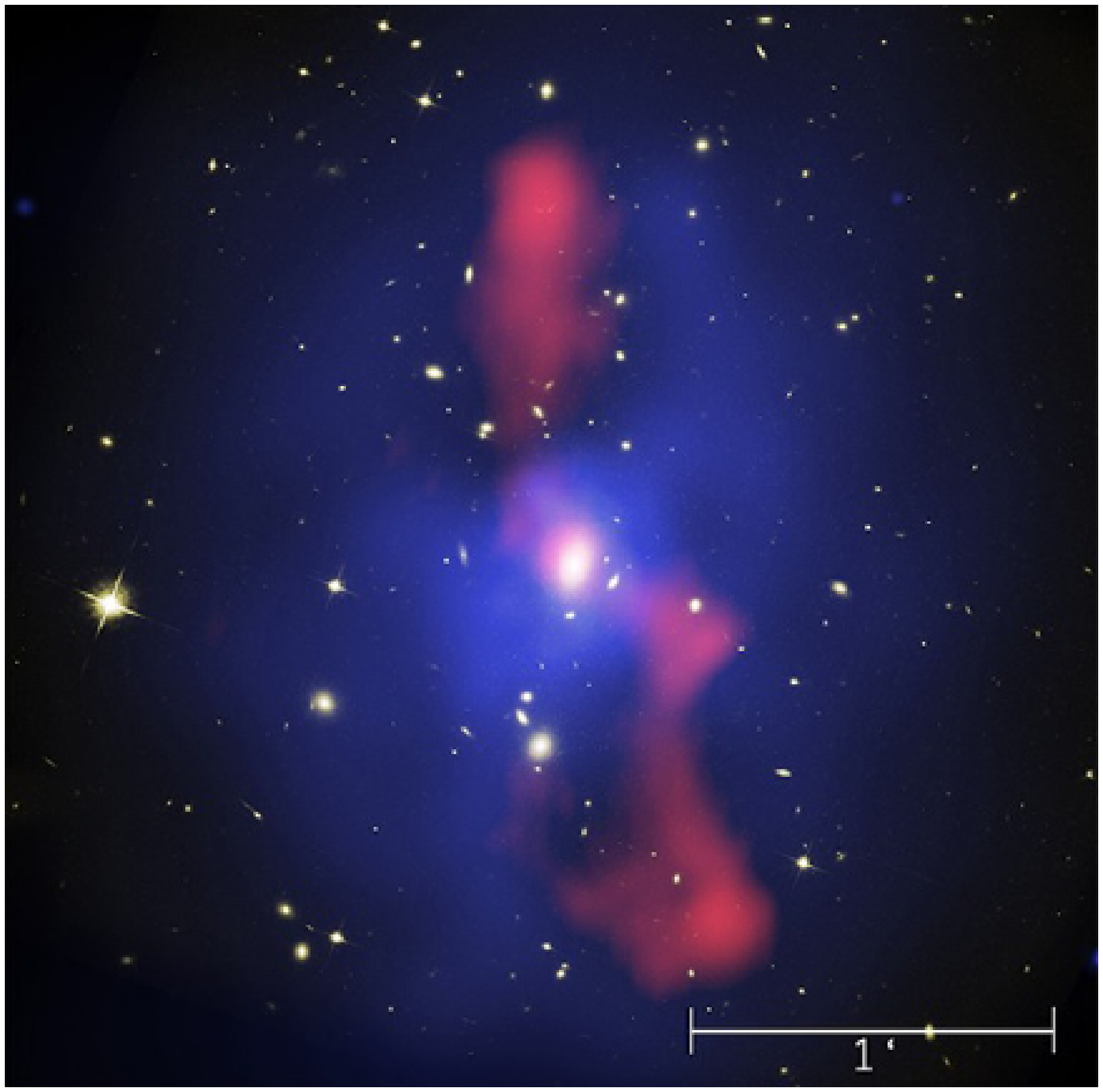}
\caption{Left: The Arms and weak shocks produced by the jets of M87
  (Forman et al 2007). Right: The gigantic interaction of the radio
  lobes and intracluster gas of MS0735.6 (McNamara et al 2009). The
  figure shows the inner 700~kpc of the cluster, extending well beyond
  its cool core.  }
   \label{fig1}
\end{center}
\end{figure}

One solution is that the tight balance is only apparent. If the jets
become too energetic then can they push through the whole cooling
region (e.g. Cyg A or MS0735.6+7421; McNamara et al 2009) and deposit
energy much further out. If cooling dominates then it can feed the
reservoir of cold gas seen in many objects, as well as some star
formation. The Brightest Cluster Galaxy (BCG) at the centre of A1835
at $z=0.25$ is an extreme example with over $100\Msunpyr$ of massive
star formation (O'Dea et al 2008). It is within a factor of two of the
highest star formation rate of any galaxy at low redshift (Arp
220). (Without heating, the central intracluster gas in A1835 would be
cooling at over $1000\Msunpyr$, so a balance remains, but not a very
tight one.)

\begin{figure}[h]
\begin{center}
 \includegraphics[width=2.2in]{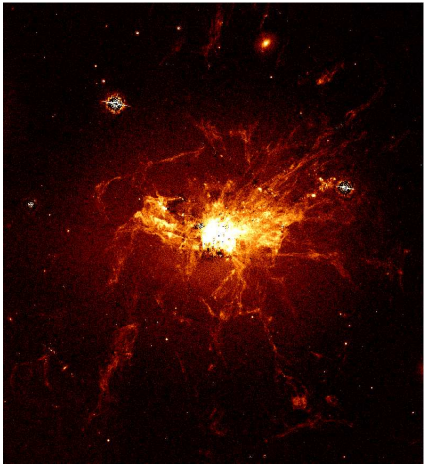}
\hspace{0.cm}
\includegraphics[width=3.0in]{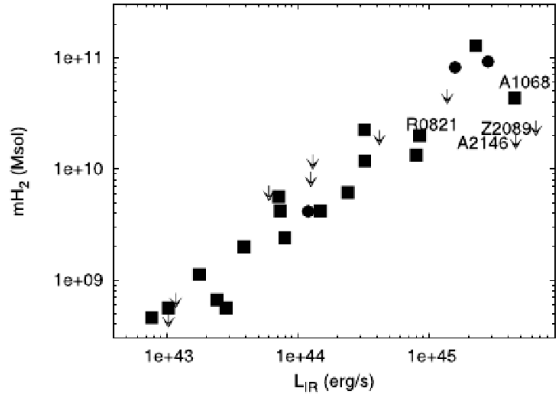}
 \caption{Left: HST image of the filaments around NGC\,1275 in the Perseus
   cluster (Fabian et al 2008). Right: Mass of H$_2$ reservoir compared with
   Spitzer IR luminosity (O'Dea et al 2008).   }
   \label{fig1}
\end{center}
\end{figure}

Many BCGs in {\it cool core} clusters (the ones with the short
radiative cooling times) have extensive optical emission-line
nebulosities (e.g. Crawford et al 1999). The cold gas is mostly
molecular as shown by CO (Edge 2001) and H$_2$ emission.  NGC\,1275 at
the centre of the Perseus cluster is a spectacular example (Fig. 6)
with its filaments being composed of about $10^{11}\Msun$ of H$_2$
(Salom\'e et al 2006). Star formation happens sporadically in that
galaxy with $\sim 20\Msunpyr$ occurring over the past $10^8\yr$ in the
SE blue loop (Canning et al 2009).  Dust is seen in the form of dust
lanes and infrared emission, with Spitzer observations revealing high
IR luminosities (Egami et al 2006, O'Dea et al 2008, Fig. 6). The dust
is presumably injected by stars into the central cold gas reservoir,
from where the bubbles drag gas out to form filaments.

Much of this IR luminosity is due to vigorous star formation in the
BCG, presumably fuelled by a residual cooling flow. Some however
could be due to the coolest X-ray emitting clumps, at 0.5--1~keV,
mixing in with the cold gas and thereby cooling non-radiatively
(Fabian et al  2002; Soker et al 2004). The outer filaments in
NGC\,1275, may be powered by the hot gas (Ferland et al 2009).

The conclusion  is that gas may be cooling from the hot phase of
the intracluster medium at a higher rate than otherwise thought. Some
of the cooling occurs non-radiatively by mixing. The gas then hangs
around for Gyrs as a reservoir of cold molecular dust clouds, 
forming stars slowly and sporadically.

Generally the central AGN in BCGs is quite sub-Eddington ($\lambda\sim
10^{-3}- 10^{-2}$). The luminous low redshift quasar H1821+643 at
$z=0.3$ is a counter-example (Russell et al 2009). The surrounding gas
however seems to be in the same state as for normal cool core BCGs.

\section{Discussion}

An active nucleus interacts with the gas in its host galaxy
through radiation pressure, winds and jets. The consequences can be
profound for the final mass of the stellar component of the galaxy as
well as  the black hole. 

The radiative or wind mode was most active when the AGN was a young
quasar. At that stage the galaxy had a large gaseous component and the
nucleus was highly obscured. Direct observational progress is therefore
difficult and slow. The kinetic mode on the other hand is more easily
observed, albeit at X-ray and radio wavelengths, since it is acting
now in nearby objects and the surrounding gas is highly ionized. An
attractive possibility is that the radiative mode shaped the overall
galaxy and black hole mass at early times and the kinetic mode has
since maintained that situation where needed (Churazov et al 2006).

\end{document}